\documentclass[11pt,a4paper]{article} 
\pdfoutput=1
\usepackage{jcappub} 
\usepackage{bm}
\usepackage{graphicx,amssymb,amstext,amsmath}
\usepackage{natbib}
\usepackage{threeparttable}
 \newcommand{\ud}{\mathrm{d}}

 \title{ The rates and time-delay distribution of multiply imaged supernovae behind lensing clusters}
 \author[a,1] {Xue Li\note{Author to whom any correspondence should be addressed.},} 
 \author [a] {Jens Hjorth}
 \author [b] {and Johan Richard}
 
 \affiliation [a]  {Dark Cosmology Centre, Niels Bohr Institute, University of Copenhagen, \\ Juliane Maries Vej 30, 2100, Copenhagen, Denmark}
  \affiliation[b] {CRAL, Observatoire de Lyon, Universit{\'e} Lyon  1,\\ 9 Avenue Ch. Andr{\'e}, 69561 Saint Genis Laval Cedex, France}
 \emailAdd{lixue@dark-cosmology.dk}

 \abstract{Time delays of gravitationally lensed sources   can  be used to constrain the mass model of a deflector and determine cosmological parameters. We here  present  an analysis  of the time-delay distribution of multiply imaged sources behind 17 strong lensing galaxy clusters  with    well-calibrated mass models.  We find  that for time delays less than 1000 days, at $z = 3.0$, their logarithmic probability   distribution functions   are well represented by    $P (\log  \Delta t) =  5.3 \times 10^{-4}  \Delta t^{ \widetilde \beta}  / M_{250} ^{2   \widetilde \beta}$, with  $ \widetilde \beta = 0.77$,   
 where $M_{250}$ is the projected cluster mass inside 250 kpc (in $10^{14} \textrm{M}_{\odot}$), and  $\widetilde \beta$  is the power-law  slope of the distribution.   The resultant   probability  distribution function  enables us to estimate the time-delay distribution in a lensing cluster of known mass.  For a cluster with   $M_{250}  = 2 \times 10^{14} \textrm{M}_{\odot}$, the fraction of time delays less than 1000 days is  approximately $  3\%$.    Taking Abell 1689 as an example,  its dark halo and brightest  galaxies,  with central velocity dispersions $\sigma \geqslant 500  \  \textrm{km} \  \textrm{s}^{-1}$,  mainly produce large time delays, while galaxy-scale mass clumps  are responsible for generating   smaller time delays.        
 We   estimate  the probability of observing   multiple images of   a supernova  in   the known images of Abell 1689.  A two-component model of estimating the supernova rate is applied in this work.
 For a magnitude threshold  of $ m_ \textrm{AB} = 26.5$, the yearly rate of     Type Ia (core-collapse) supernovae with time delays  less than 1000 days  
is $0.004 \pm 0.002$ ($0.029 \pm 0.001$).    
If the magnitude threshold is lowered to   $ m_ \textrm{AB} \sim 27.0$,   the rate of   core-collapse  supernovae  suitable for time delay observation is    $0.044 \pm 0.015$  per year.  }  
 
 
\begin{document}
 \maketitle
\flushbottom

\section{Introduction}

   An object in our universe, such as  a galaxy or a galaxy cluster,   could   bend light rays passing  it and act as a lens  to magnify or demagnify sources behind it \cite{zwicky}. This effect is known as   gravitational lensing and has been developed into a powerful cosmological tool in recent decades  \cite{Cooke_1975} \cite{Blandford_1992} \cite{Efstathiou_1999}  \cite{Kneib_2011}.   
  With the help of gravitational lensing, we can observe   distant galaxies   behind   galaxy clusters  which would otherwise be too faint to be   observed, and analyze their properties. We can also measure the cosmological parameters that describe the  geometry and  the expansion rate of the universe       \cite{Refsdal_1964} \cite{julloscience2010}. In addition, we can   analyze  the total  mass distribution in lensing galaxy clusters \cite{kochanek_2004},    regardless of the differences between  luminous   and dark matter \cite{Massey_2010}.


In strong gravitational lensing systems, multiple images are produced. 
Light travels along the stationary paths between two points in space time. A  massive   object,  like a galaxy or cluster of galaxies located along the light path,  in general affects and perturbs the light trajectory \cite{Schneider}.  
When lensed by a   galaxy or a cluster of galaxies,  light  emitted by a source  may  travel  along   different light paths and be observed as different multiple images.    Light from these images is received at different times. Thus, multiple images have different arrival times  to the observer.  The difference of arrival times between multiple images of the same source   is called the  time delay.

So far,     time delays     have been studied and applied    in many ways:  to     constrain  the Hubble parameter $H_{0}$ 
 \cite{Saha_2006}    \cite{danuta_p} \cite{Suyu_2012};
   to study the galaxy mass profile with   Monte Carlo simulations  \cite{rusin};  to measure the cosmological parameter   $w$  \cite{danuta} \cite{dancoe} \cite {suyu} \cite{Linder_2011} \cite{Suyu_2012}; 
   to improve the mass models of     galaxies  with the time delays of  quasars \cite{luis}.

Compared to quasars,  light curves of type Ia supernovae   (SNe Ia)  evolve regularly with time, and have been extensively studied \cite{Hamuy_1996} \cite{Hillebrandt_2000}.  Hence, they  are potentially very useful as standard sources for constraining   time delays in gravitational lens systems.  In principle,  they   can also be used to constrain the Hubble constant  through the  measurement of  the     time delays \cite{Oguri_2003}.  
SNe Ia  play a key role as   standard candles   in distance measurements  on cosmological scales.  
Supernovae   provide direct evidence that the low-redshift universe is accelerating \cite {riess} \cite{perlmutter} \cite{nordin}.  They   act as the primary sources of heavy elements and  potentially dust   in the universe \cite{Gall_2011}.  
However, there is still debate on the progenitor models of SNe Ia \cite{Hillebrandt_2000} \cite{Maoz_review}. 
There are two major progenitor scenarios to explain the mechanism of Type Ia progenitor.  In the single degenerate model \cite{Whelan_1973},  a carbon-oxygen white dwarf    accretes  mass  from a companion star, a subgiant, a helium star or a red giant, and reaches the Chandrasekhar mass limit, resulting in a thermonuclear explosion \cite {hoyle} \cite{branch}. In the double degenerate scenario \cite{Webbink_1984}, two white dwarfs merge,  approach the  Chandrasekhar mass limit, and ignite. 
Recently, a third model gives another explanation of the possible SN progenitor scenarios. Instead of accreting the mass to the  Chandrasekhar mass limit, the detonation ignites to the accreted He shell of one white dwarf, then the detonation shock wave comes to the core or near the center, and a second detonation happens \cite{Fink_2007} \cite{Bildsten_2007} \cite{Guillochon_2010}. 
 
The rate of supernovae  (SNR) reflect their formation mechanism.   For example, core-collapse SNe (Type II and Ibc supernovae) arising from massive stars help us trace the star formation and 
 may  be used   to constrain the star formation rate (SFR) \cite{Anderson_2011}. 
 A well established  model  of estimating the SNR for Ia SNe ($\textrm{SNR}_\textrm{Ia}$)   is  a 	\textquotedblleft two-component\textquotedblright \  model \cite{Scannapieco_2005}        \cite{Mannucci_2005}  \cite{Sullivan_2006}, with one     component  dependent on the recent star formation in the host galaxy and the other   component   dependent on the host stellar mass. 
  The SN type Ia rate is a combination of these two components.  
 The $\textrm{SNR}_\textrm{Ia}$ at intermediate redshift has been constrained using the SDSS-II dataset to $ z \leqslant 0.12$ \cite {Dilday_2008}, and extended analysis to  $ z < 0.3$  \cite{Dilday_2010}. At high redshift, the SN Ia rate is also tested and constrained, using the SNLS dataset to $ z \thickapprox 0.5$ \cite{Neill_2006} and $ 0.1 \le z \le 1.1$ \cite{Perrett_2012},   the   {\it HST}/GOODS survey up to $ z < 1.8$ \cite {dahlen_19} \cite {goods_2},  
 and the Subaru Deep Field (SDF) to  $z < 2$ \cite{Graur_2011}.   The core-collapse supernova  rate is also tested and estimated  up to $z \sim 0.7$, using the GOODS survey. With these data and the SNR model, we can estimate the lensed SNR in a cluster lensing system \cite{Riehm}. 

The aim of this paper is to  (1)  develop a general function to describe   the time-delay distributions in gravitational lensing systems;  
(2) estimate the lensed SNR as a function of time delay and magnitude in Abell 1689  to assess the feasibility of constraining mass models and cosmological parameters with lensed supernovae observationally. 

The outline of the 
 paper is as follows.  In section \ref{sec:basicequation},  we  develop a theoretical formalism for describing the time-delay distribution.   The   analysis  and discussion of  parameters   for the  distribution function are developed in section  \ref{sec:abell1689}. In section \ref{sec:17clusters},   we   model  time-delay distributions  of 17 massive lensing clusters.  We analyze and fit the parameters of the  distribution function, based on the results    from modeling.     In section  \ref{sec:snr},  using the  \textquotedblleft two-component\textquotedblright \  model, 
 we calculate the probability of observing   supernovae in   35 known multiply imaged galaxies behind   Abell 1689.   Finally, we summarize our investigation and discuss future prospects in section \ref{sec:Summaryanddiscussion}. Throughout this paper, we assume a  cosmological model with $ \Omega_{m} = 0.3,  \Omega_{\Lambda} = 0.7, h = 0.7$. Magnitudes are in   the {\it AB } system.

\section{Time delay theory}
\label{sec:basicequation}

A light ray is deflected    when it passes   a cosmic   massive object.   In a  lensing system,   the   light   path  from the source to the observer is changed according to the   gravitational field near the  lens.   In the case of a multiple image system,  lensing generally  causes a difference in the arrival time  of a galaxy image pair and hence generates a time delay.   The time delay, $\Delta t$,   can be calculated as   \cite{Schneider}

    \begin{equation} 
c \Delta t (y) = \xi ^{2}_{0} \frac{D_\textrm{OS}}{D_ \textrm{OL}D_ \textrm{LS}}(1+z_{L})[\phi(x_{1},y)-\phi(x_{2},y)],
  \label{timedelaytheory}
\end{equation}  
where  $\xi_{0} = 4 \pi (\frac{\sigma_v}{c})^2 \frac{D_ \textrm{OL}D_ \textrm{LS}}{D_\textrm{OS}}$    
is the characteristic length scale in the lens plane.    
Here $\sigma_v $ is the value of an effective velocity dispersion,  $D_\textrm{OS} $ is the angular diameter distance between the observer and the source,   $D_ \textrm{OL} $ is the angular diameter distance between the observer and the lens,   $D_ \textrm{LS} $ is the angular diameter distance between the lens and the source,  and   $z_{L}$ is the lens redshift.  
We denote the image position in the lens plane by $ \bm{\xi}$ and the source position in the source plane by $ \bm{\eta}$.
   Here $y =  | \bm{\eta}| / \eta_{0}$,   is  the source position in  the source plane, with  $\eta_{0} = \frac{\xi_{0} D_\textrm{OS}}{D_ \textrm{OL}}$   being the maximal distance to the caustic line. 
We define $x_{i}  (i = 1,2)  $  as    two image positions  in the lens plane  with  $x =  |\bm{\xi}| / \xi_{0}$. Here $x, y$ are dimensionless vectors.   The Fermat potential is defined as $\phi (x_{i},y) (i = 1,2)$. For a two-image system,  the larger the difference of their Fermat potentials, the larger    time delay they will generate.  The Fermat potential $\phi (x ,y)$ can also be  described by the lensing potential $\varphi (x)$ \cite{Narayan},
    \begin{equation} 
\phi(x,y)=\frac{(x-y)^{2}}{2}-\varphi (x).
\end{equation}
%
 We assume spherically symmetric lenses in what follows.

 Generally, for a lens with density distribution    $\rho \varpropto r^{- \delta}$,  the time delay can   be expanded as \cite{kochanek_2004}  \cite{witt_1}   
     \begin{equation}  
 \Delta t ( \delta) \approx (\delta -1) \Delta t_{\textrm{SIS}} \left[ 1 - \frac{(2- \delta)^2}{12}   \left(\frac{\Delta r}{ \langle r \rangle }\right)^2 \cdots \right], 
  \label{internal}
\end{equation}
 where $\Delta t_{\textrm{SIS}}$ represents the time delay  for the Singular Isothermal Sphere   (SIS; $\delta = 2$),   and $ \langle r \rangle = (r_i + r_j)/2$.    If the term $\frac{\Delta r}{ \langle r \rangle }$ is small, the higher   order terms can be ignored.  
 
  For  real clusters,    tidal perturbations from   objects near   the lens or along the line of sight \cite{Keeton_1997} \cite{Witt_1997} may affect the images as well.   An external shear     can be added to the lens  \cite{witt_1}, whose potential  is   
     \begin{equation} 
 \phi(\gamma) = - \frac{1}{2}[ \gamma_1  (\zeta^2_1 - \zeta^2_2) + 2  \gamma_2 \zeta_1 \zeta_2],   
\end{equation}
 where $\gamma$ is the strength of the shear,   $\gamma_1 = \gamma \cos 2 \theta_\gamma$ and   $\gamma_2 = \gamma \sin 2 \theta_\gamma$, with  $\theta_\gamma$ being  the angle between the direction of the shear and the   major axis of the lens.  Here $ \zeta_1$,  $ \zeta_2$ represent the  components of the lens coordinate.   Then   more than 2 multiple images may be produced by each source, and the time delay between   images $i$ and $j$ is \cite{witt_1}  
     \begin{equation}  
 c  \Delta t   =  \frac{D_\textrm{OL} D_\textrm{OS}}{2  D_\textrm{LS}} (1+ z_L) \{(r^2_j - r^2_i) + \gamma [  r^2_j \cos 2(\theta_j - \theta_\gamma) - r^2_i \cos 2 (\theta_i - \theta_\gamma)         ]          \}, 
\label{external}
\end{equation}
  where $r_k = ( \zeta_k^2 +  \zeta_k^2)^{1/2}$ with $k = i,j$ is the distance of the image  from the center.    The shear   affects     the time delay in   two-image lenses with $r_i \ne r_j$ only slightly, while in   four-image lenses with $r_i \approx r_j$, the shear may play a significant role.   

 For a lens in a general quadrupole with total shear $\Gamma = \gamma_{int} + \gamma_{ext}$, the time delay  can be estimated as    \cite{kochanek_2004} 
\cite{Kochanek_2002} 
  \begin{equation}  
\Delta t  \simeq 2 \Delta t_{\textrm{SIS}}    (1 -\langle \kappa \rangle)    \frac{\sin^2 (\Delta \theta_{ij}/2)}{1-4 f_{int}\cos^2 (\Delta \theta_{ij} /2)}                 ,  
\label{bothshear}
\end{equation}
 where $\langle \kappa \rangle$ is the average surface density in the annulus bounded by the images in units of the critical surface density and $  f_{int} =  \gamma_{int} /\Gamma$ is the fraction of the quadrupole.    Here $\Delta \theta_{ij}$ represents the angle between the images  $\Delta \theta_{ij} = \theta_1 - \theta_2$.

   To discuss the time-delay distribution in a lensing cluster, we start from a simple situation, in which the mass distribution of a lensing cluster is described by a     SIS  profile and no more than two multiple images are generated from each source. With   the  density   $\rho (r) = \frac{\sigma_v^2}{4 \pi G r^2}$,   
 the time delay is \cite{oguri} 
  \begin{equation}
c \Delta t_{\textrm{SIS}} =    32\pi^{2}\left( \frac{\sigma_{v}}{c} \right)^{4}\frac{D_ \textrm{OL}D_ \textrm{LS}}{D_\textrm{OS}}(1+z_{L})y,         
 \label{sis}
\end{equation} 
%
from which  it follows that  $\Delta t \propto y$, i.e., the time delay is proportional to the  location of the source.  If the source is located within the strong lensing area enclosed by the caustic line,  two images will be created. In this case,  the normalized probability  that the source     is located between $\eta$ and $\eta + \ud \eta$ is \cite{samsing_private2010} 
 \begin{equation} 
 p  (\eta,   \eta+\ud \eta) = \int _{\eta}  ^{\eta+ \ud \eta} 2\frac{\eta}{\eta ^{2}_{0}} \ud  \eta ;  \qquad  \eta \leqslant \eta_{0}.
    \label{theory_r}      
\end{equation}  
Hence,  the normalized time delay probability   distribution function can be simplified as  
   \begin{equation} 
    \label{theory}
f (\Delta t ) = 2\frac{ \Delta t }{\Delta t ^{2}_ \textrm{peak}} ;     \qquad       \Delta t \leqslant     \Delta t_ \textrm{peak}.             
\end{equation} 
Here $f(x)$ denotes a  probability   distribution function throughout this paper and $\Delta t_ \textrm{peak}$ is the time delay which has the largest   probability. 
With 
 $f (\Delta t ) \ud (\Delta t ) =   P (\log \Delta t ) \ud (\log \Delta t )$,  and $ \ud (\Delta t )  = \ln 10 \cdot 10^{ \log  \Delta t} d (\log \Delta t )$,  
  it follows that $P (\log \Delta t ) = \ln 10  \cdot 10^{ \log  \Delta t}   f  (\Delta t )$, i.e.,  the  probability   distribution function  of the logarithmic time delay  for the SIS profile is
   \begin{equation} 
P (\log \Delta t ) = \frac{2 \ln 10}{\Delta t ^{2}_ \textrm{peak}} 10^{2 \log\Delta t };     \qquad       \log \Delta t \leqslant     \log \Delta t_ \textrm{peak}.          
    \label{theory_2}   
\end{equation} 
    The time-delay distribution is   sensitive to  the slope  of the density profile \cite {Keeton_2001} \cite{Wyithe_2001}.  Steeper inner slopes tend to produce larger time
delays \cite{oguri}.  There are several theoretical profiles  established  to  describe the mass distribution of a cluster: the SIS  profile, the dPIE profile (dual pseudo isothermal elliptical profile) \cite{ardis_mass}, and the NFW profile (Navarro-Frenk-White profile) \cite{navarro}, etc. 
For a density distribution   described as    $\rho \varpropto r ^{- \delta}$,  the    density slopes $\delta$   for these three  profiles are listed in table \ref{densityprofiles}.  
The slope of the density profile may affect the distribution of the  time delay in this way: compared to the SIS profile, the NFW profile has shallower inner density slope, but steeper outer density slope,  which means that  on small time delay scales,  the distribution of the time delay will be stretched out to a higher
probability,  but on   large time-delay scales, it will be lower than the distribution of a SIS profile. The  dPIE profile has even shallower inner density slope but steeper outer density slope than the NFW profile, so  the time-delay distribution of dPIE profile has the shallowest slope among the three profiles. 
\begin{table}

\centering
\begin{tabular}{|c c c c c|}

\hline          
    &              inner $\delta$  &        intermediate $\delta$   &        outer $\delta$ &         $\beta$ \\ 
  \hline

 SIS & $-2$ &   $-2$  & $-2$ & $2.00$ \\
 NFW & $-1$&   $-2$   & $-3$ & $1.20$\\
 dPIE & $0$ &  $-2$  & $-4$ & $0.83$ \\

 \hline
 \end{tabular}
 \caption{ The slopes of the density profiles describing the mass distribution of a galaxy cluster and the slopes of the best-fitting functions to the time-delay distribution. More details of the best-fitting function and slope $\beta$ are described in   subsection \ref{subsec:betaanalys}.  } 
 \label{densityprofiles}
\end{table}  
Equations (\ref{theory}) (\ref{theory_2})   show  that the time-delay probability   distribution function is  a power law  for the SIS profile.   
Motivated by the  discussion above, we make the  simple ansatz that    the time-delay probability   distribution function can be approximated as:
  $f (  \Delta t)    \propto   \Delta t ^{\beta-1}$ and  $P (\log  \Delta t)  \propto   10^{ \beta \log \Delta t}$, or $   P (\log  \Delta t)  \propto   \Delta t ^{\beta }$. More details on the comparison of the ansatz and   simulation results are  presented   in section  \ref{subsec:betaanalys}.

 This ansatz is based on the assumption of the time-delay  distribution generated by a two-multiple-image system.  
  For  clusters with complicated structures, as shown in equations (\ref{internal}), (\ref{external}),  (\ref{bothshear}),  e.g., clusters whose mass distributions  are described by clumps of potentials, the situation is more complicated because more images are produced so more time delays are generated. Fortunately,  the simulations described in section \ref{subsec:massanalys}  and figure \ref{abell1689function_1}  show that  the  distribution of   time delays  generated by multiple images still obey a power-law distribution. 
  
   After normalization, the probability   distribution functions can be written as:
 \begin{equation} 
    \label{ansatz_2}
f (  \Delta t) =     \frac{\beta }{ \Delta t ^{\beta}_ \textrm{peak}}     \Delta t ^{\beta - 1};                \qquad             \Delta t  \leqslant   \Delta t_ \textrm{peak},
\end{equation}    
 \begin{equation} 
    \label{ansatz_1}
P ( \log  \Delta t) =     \frac{\beta \ln 10}{ \Delta t ^{\beta}_ \textrm{peak}}     \Delta t ^{\beta};                  \qquad             \Delta t   \leqslant  \Delta t_ \textrm{peak},
\end{equation} 
 \begin{equation} 
    \label{ansatz_3}
\log P ( \log  \Delta t) =   \beta \log ( \Delta t )  +  \log ( \beta \ln 10 ) -  \beta  \log \Delta t_ \textrm{peak};                \qquad             \Delta t   \leqslant  \Delta t_ \textrm{peak}.
\end{equation} 
We can also write the cumulative probability distribution as 
\begin{equation} 
    \label{cansatz_1}
  f ( <  \Delta t)  =  \int_{0}^{ \Delta t  }  f (x) d x =  \left \{    \begin{array}{ll}
   \frac{\Delta t ^{\beta} }{ \Delta t ^{\beta}_ \textrm{peak}}                &     \qquad     \Delta t   \leqslant  \Delta t_ \textrm{peak},     \\
                                                         1                                            &    \qquad   \Delta t  >  \Delta t_ \textrm{peak}. \end{array}
 \right.
\end{equation} 
If we set     $\beta = 2$, the   probability   distribution functions (\ref{ansatz_2}) and  (\ref{ansatz_1}) reduce  to the expressions for the SIS case, i.e.,  functions (\ref{theory})   and  (\ref{theory_2}).

According to the virial theorem\cite{james}, for a fixed radius, the enclosed mass is   proportional to the velocity dispersion, 
$  M   \varpropto \sigma_{v} ^2$.   From equation (\ref{sis}),  for the SIS profile,  $   \Delta t_\textrm{peak, SIS} \varpropto   \sigma^{4}_{v} \varpropto  M ^{2}$.  
More generally, for a  general profile  with density distribution   described as    $\rho \varpropto r ^{- \delta}$, 
 the expression for the time delay can be well approximated  as  
    $ \Delta t \varpropto \Delta t_\textrm{SIS} (\delta-1) $ (\ref{internal}),   
i.e.,    $\Delta t \varpropto  (\delta -1 ) M^{2}$. 
Hence, we can    write the probability   distribution function (\ref{ansatz_3})  as
   \begin{equation} 
    \label{function_betac3}
\log P ( \log  \Delta t) =   \beta \log ( \Delta t )  +  \log ( \beta^{C_{1}} \ln 10 ) -  2 \beta  \log M_{250}  + C_{2},
\end{equation} 
where   $C_{1},C_{2} $ are constants to be determined.  
Here   $M_{250}$ is defined as the  projected mass within $R <250  \, \textrm{kpc}$, in units of $10^{14}\, \textrm{M}_{\odot}$.  
  In this equation, $\beta$ and   $M_{250}$ are parameters. 
  If the value of $\beta$ is fixed, on the right hand of the equation (\ref{function_betac3}), the second  and the  fourth terms (i.e., $\log ( \beta^{C_{1}} \ln 10 )$ and $C_{2}$) will be reduced to one constant: $ C^{'}_{2} =  C_{2} +   \log ( \beta^{C_{1}}_{\textrm{fix}} \ln 10)$. In this case, there will be only one constant $ C^{'}_{2}$ to be determined.  So the logarithmic probability distribution function   can be reduced to
   \begin{equation} 
    \label{function_betac}
\log P ( \log  \Delta t) =   \beta \log ( \Delta t )   - 2 \beta  \log M_{250} + C^{'}_{2}.
\end{equation} 
We will discuss   $\beta$ and   $ C^{'}_{2}$ in section \ref{sec:abell1689}.  

  
 \section{Modeling clusters of galaxies}
\label{sec:abell1689}   

 In this section, we   discuss   how $\beta $  depends on different mass profiles,  and the effect of the mass on the time-delay distribution.  Considering the time required for a realistic observation,    small time delays, i.e., time delays no longer than 1000 days, are more suitable for an actual time-delay measurement in a reasonable amount of time. In this paper, we  therefore focus on time delays less than 1000 days.    We  treat all time delays as independent from each other   and the time delays are taken from all images pairs.    

   \begin{figure}

 \centering
\includegraphics [scale = 1, bb=0 0 800 283] {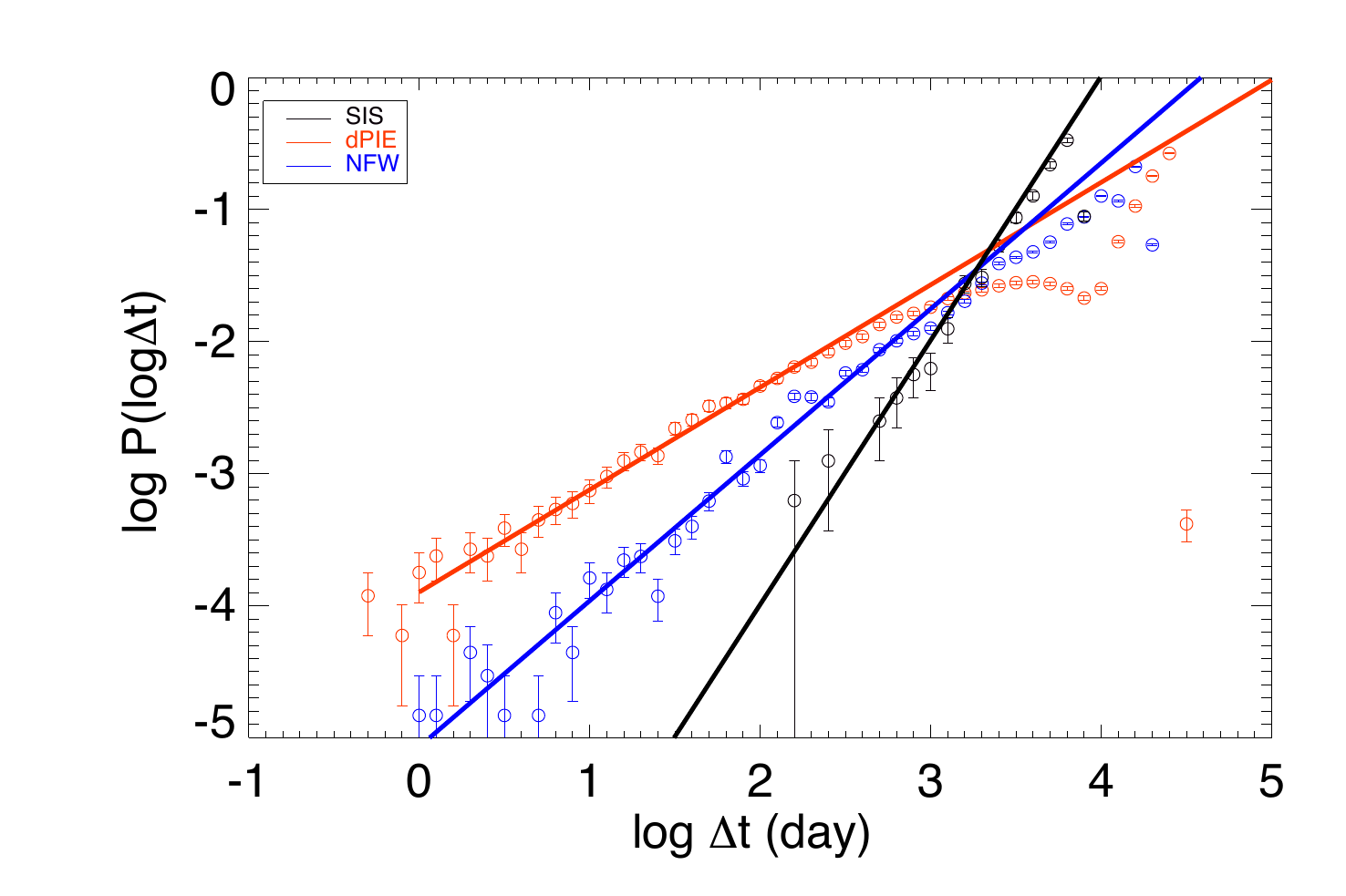}

\caption{  The  logarithmic probability   distribution    of the time delays for three density profiles.  They are   the SIS profile (black), the dPIE density profile  (red) and the NFW density profile (blue).  The    background sources are at $z=3$.    The open circles and their poisson error bars are obtained  from a  numerical representation of the potentials.       The best-fitting  functions for each cluster are also plotted.    The slopes of the best-fitting functions for the SIS profile, the NFW  profile and the dPIE  profile are 
$ \beta_{\textrm{SIS}} = 2.0$, $\beta_{\textrm{NFW}} = 1.11$, $\beta_{\textrm{dPIE}} = 0.78$.  }
  \label{theorysispiemdnfw}
\end{figure}

 To get the time-delay distribution, we   create an input catalog of sources.  
Time delays  are created when there are multiple images,  so we need to make sure that the input source plane covers the area enclosed by the caustic line(s) and includes all  potential multiply lensed sources.  On the other hand, the input source plane should be sufficiently well sampled so as to be sensitive to the mass distribution and potential differences in the lens.  This is to make sure that small time delays are also  produced.  In this work, we choose an  input source plane covering an area of 60 $\times$  60 arcsec   with 200 $\times$ 200 pixels in the source plane at $z = 3$. 
With the help of  {\it Lenstool} \cite{Jullo_2007}, we  can obtain  a    list of   all   images of every source in the source plane. Then we  compute the differences in  the arrival times   between  multiple  images from each source.   
 For example, for a source with 3 multiple images: image1, image2, image3, we compute the differences of arrival times between each two of these three images. Then we get 3 time delays in total.

\subsection{The slope of the time-delay distribution}
\label{subsec:betaanalys}

The time-delay distributions for the SIS,  the NFW and  the  dPIE  profiles   are simulated and computed.    
In simulating the time-delay distributions, we choose parameters similar to those of mass models of real clusters. We adjust the parameters of the profiles  to make their distributions overlap. The results are shown    in   figure \ref{theorysispiemdnfw}.   In the NFW mass model, the velocity dispersion is 1810  km/s,  the concentration is  3.579 and the scale radius is 618 kpc.    In the dPIE mass model, the velocity dispersion is 1500 km/s, the core radius is 72 kpc,  and the cut radius is 2000 kpc.   For the mass model described by the SIS profile, we set  the velocity dispersion   to 400 km/s.   A change in the velocity dispersion will shift the distribution  horizontally but keep the slope unchanged.    So  for   computational convenience, we keep the velocity dispersion and    shift the  distribution  horizontally  to make  the   results  for three profiles overlap.     All three profiles are circular.   With   time delays  less than 1000 days, the   slopes  of  the logarithmic probability   distribution functions for the SIS profile,  the NFW  profile and  the dPIE  profile,  are $ \beta_{\textrm{SIS}} = 2.0$ as expected, $\beta_{\textrm{NFW}} = 1.11$, $\beta_{\textrm{dPIE}} = 0.78$, respectively (see also table \ref{densityprofiles}). 
The difference in slopes of the distributions  arises from  the difference in the  density slopes of  the mass profiles, as discussed in  section \ref{abell1689function_1}.   This shows that the slope of the time-delay distribution is strongly affected by the mass distribution, especially the density slope of the cluster.

\subsection{An example: Abell 1689}
  \label{subsec:massanalys}


  \begin{figure} 

  \includegraphics    [ scale = 1] {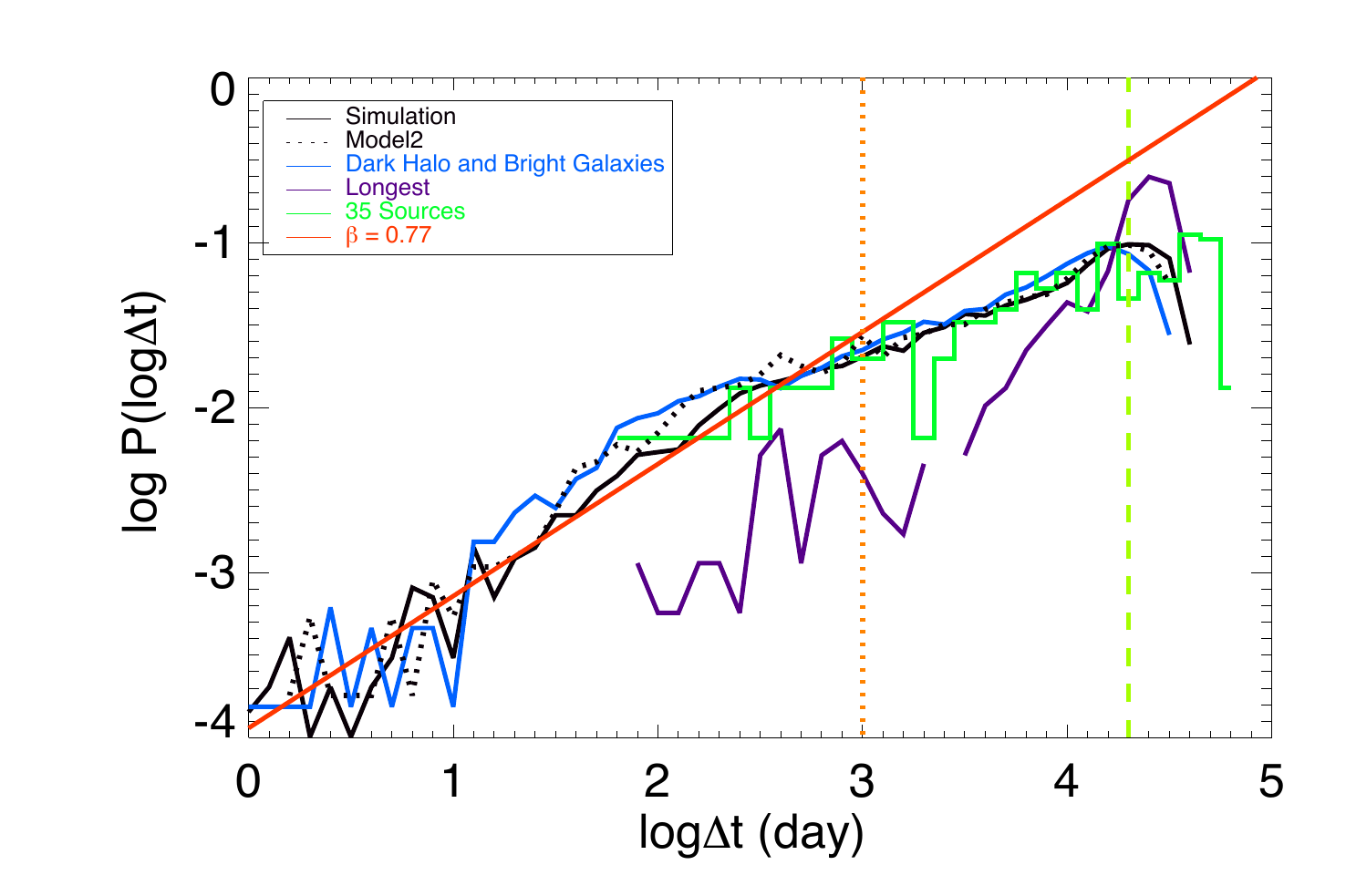}
\caption{  The  time-delay   distributions of Abell 1689 with  modeled  background sources  at $z=3$. 
The   time-delay  distribution produced by the   cluster  with the full mass model is in black. The  blue curve represents the time-delay distribution produced by  the dark halo and a few bright  galaxies (massive components), which fails to reproduce small  time delays.   The distribution of the selected longest time delays from each source  is in purple.   The red curve represents the logarithmic probability   distribution function with $ \beta  = 0.77$.  The histogram in green  represents the distribution of the time delays of  35 known multiply imaged sources.      We are mainly  interested in small  time delays, i.e., those located to the left   of  the dotted orange line at 1000 days ($\log \Delta t $ = 3.0).  
The dashed line  in  olive  represents   $\Delta t_ \textrm{peak}$.  
}
  \label{abell1689function_1}
\end{figure}
With the help of  deep Hubble Space Telescope ({\it HST}) imaging,   Abell 1689 ($z = 0.183$) displays a large  number of multiple image systems   at the center of the cluster.
  Using information from these  systems, the mass model of Abell 1689 has been extensively studied.  
We base our work on the mass model consisting of  35 plausible  lensed sources and 116 multiple images \cite{MarceauLimousin} \cite{richard_private2011} \cite{julloscience2010}.   Among them, 25 sources   have confirmed spectroscopic redshifts, the other 10 systems have lensing modeling redshifts  or/and photometric redshifts.  

To check how the time-delay distribution   depends on the mass distribution, we split the mass model of Abell 1689 into two parts. 
The first part (massive  components)   consists of  a dark halo   and the three brightest   galaxies with central velocity dispersions $\sigma \geqslant 500 \,  \textrm{km} \, \textrm{s}^{-1}$. The second part  includes all other substructures (subcomponent), that is,  all other   gravitational potentials with  velocity dispersions   $\sigma$ smaller than $ 500 \, \textrm{km} \, \textrm{s}^{-1}$.
 We compute the time-delay distribution  in two different ways:
First,  we produce all time delays for Abell 1689 with the full mass model. Second,  we compute time delays of Abell 1689 with only the mass model of  the  first   part (massive components).  
The second   part  (subcomponent) itself produces only two multiple images, i.e.,  only one time delay. This is because most substructures   do not gain enough mass to surpass the  critical surface  mass density, which is required for the structure to produce multiple images.

The time-delay distribution   for Abell 1689 is shown  in figure \ref{abell1689function_1}.   The histogram in green represents the distribution of the time delays of  the 35  known sources.  It  is consistent with the simulation of the grid of hypothetical sources.     A  logarithmic probability   distribution function  (\ref{ansatz_3}) with slope $  \beta  =  0.77$  in red is also plotted.  The blue curve represents the time delay distribution generated by the first part (massive components). 
 From the figure, it is evident that the first part (massive components)  succeeds  in reproducing most relatively `large' time delays, e.g., time delays larger than around 30 days. 
We conclude that small time delays are predominantly generated by substructures  in the mass model.

 The typical offset between   the observed and modeled images  is about  $\sim$ 1$''$ \cite{Kneib_2011}.        Therefore, the individual time delays are affected.   
To test how the position offset affects the time-delay distribution, we   plot the time delay distributions  generated by two different mass models for Abell 1689. 
The result is shown in figure \ref{abell1689function_1}. The black curve represents the time-delay distribution applied in this paper, while the dotted curve is the result when another mass model \cite{richard_private2010}  is used.    Though different models may produce different time delays, the slope of  the distribution of the time delays is not strongly affected.

 The time-delay distribution of a real cluster may be more complicated  than  the ones discussed in section \ref{sec:basicequation}. The structure of the lensing cluster may   consist  of more clumps of potentials with more complicated mass distribution, thus  each source may produce more than 2 multiple  images. To check how these extra multiple images affect the time-delay distribution and whether the  distributions for real clusters can also be fitted with  the power-law functions,  
we also plot   the distribution of the longest time delays generated by each source     in figure \ref{abell1689function_1}.   
The figure shows firstly that compared to the `longest' time delays,   
  the total time-delay distribution can be fitted to a power-law function. Secondly,  for small time delays, though none of  them belong to the   `longest'  time delays and there is no direct connection between the ansatz motivated by  two   image  systems and those   non-longest time delays,  the distribution of the small time delays may also be fitted with a power-law function, which is implied by the simulation result of the total time delays. 
%
%
%
Finally, even though more  multiple images  are generated and more time delays are produced, the time-delay distribution still can be fitted well with  a  power-law function.  Therefore, we proceed to apply the ansatz of the power-law functions (\ref{ansatz_2})  (\ref{ansatz_1}) deduced from two-image system to the real clusters with more multiple images.

\section{Time delays in 17 clusters }

\label{sec:17clusters}

\subsection{Cluster selection and modeling}

\begin{table}

\centering

\begin{tabular}{|   l   c r@{.}l r@{.}l  c c|}

\hline  Cluster & $ z  $ & \multicolumn{2}{c}{ RA  (J2000.0)}  &\multicolumn{2}{c}{ Dec (J2000.0)}      & $M_{250}$ &    $\beta$ \\
&   &    \multicolumn{2}{c}{  [deg] } &  \multicolumn{2}{c}{  [deg] }       &($10^{14} \textrm{M}_{\odot}$)  &\\
  \hline

Abell 2204 &0.152   &248&195540& 5&575825&     $  2.29  \pm 0.50$ &   $  0.79\pm 0.03$ \\

Abell 868& 0.154&146&359960& $-$8&651994  &   $1.97\pm 1.11$ &    $ 0.57 \pm 0.03$\\
  
RXJ 1720 & 0.164   &260&041860 & 26&625627 &       $1.18  \pm 0.59 $ &$     0.64 \pm 0.04$\\

 Abell 2218 &0.171  &   248&954604 & 66&212242&    $  3.00 \pm 0.24 $ &  $   0.89 \pm 0.04$\\
 
 Abell 1689 &0.183  & 197&872954& $-$1&341006 &      $ 4.53  \pm 0.13$  &   $  0.79 \pm0.03$\\
 
 Abell 383 &0.188 &42&014079 & $-$3&529040  &        $1.87\pm 0.26$ &  $   0.85 \pm 0.09$\\
 
 Abell 773  &0.217 & 139&472660& 51&727024 &       $3.01\pm 0.58$ &   $  0.86 \pm 0.04$\\

RXJ 2129 & 0.235   &  322&416510 & 0&089227 & $ 1.37 \pm 0.37$ & $    0.74 \pm 0.03$\\
 
Abell 1835&0.253 & 210&258650 &2&878470  &      $ 2.83 \pm 0.41$ &  $   0.89\pm 0.05$\\

Abell 1703 &0.280   & 198&771971 & 51&817494 &       $ 2.98 \pm 0.09$ & $    0.77\pm 0.08$\\

MACS 2135& 0.325  &323&800390 &$-$1&049624&      $2.64 \pm 0.04$ &  $   0.83 \pm 0.02$\\
MACS 1319& 0.328  &200&034880& 70&077501& $2.28 \pm 0.26$ & $    0.41 \pm 0.06$\\
MACS 0712&0.328   &108&085460 &59&538994 & $1.29 \pm 0.27$ &$ 0.63 \pm 0.05 $\\
MACS 0947 &0.345   &146&803230 &76&387101&$2.96 \pm 0.94$&$ 0.59 \pm 0.09$\\
 SMACS 2248&  0.348 & 342&183260 &$-$44&530966 &     $2.87 \pm  0.06$  & $    0.80 \pm 0.06$\\

MACS 1133&0.389   &173&304880 &50&144436 &$1.52 \pm 0.23$ & $     0.91 \pm 0.05$\\

 MACS 1347 & 0.451   &  206&877570 &$-$11&752643&     $3.86 \pm 0.02$ &  $   0.77 \pm 0.05$\\

 \hline
 \end{tabular}
 \caption{ Properties of 17 lensing clusters. The redshifts of the selected clusters range from 0.15 to 0.30 for models from LoCuSS and from 0.30 to 0.45 for MACS clusters.    Here $M_{250}$  denotes the projected lensing mass inside 250 kpc  \cite{richard} \cite{richard_private2010}.   
The  slopes of  $\beta$  in function (\ref{function_betac}) are also listed.  }
 \label{clusterreference}
\end{table}

To further analyze the logarithmic  probability   distribution function of time delays  and constrain the   parameter and the constant  in equation (\ref{ansatz_3}), we compute time-delay distributions by modeling 17 lensing clusters. The cluster selection procedure is based on two criteria:  First, each   lensing cluster system should have at least one image with a spectroscopically-confirmed redshift.  Furthermore, the range of redshifts of the selected clusters  should be as large as possible.  The redshifts of the selected clusters range from 0.15 to 0.30 for models from LoCuSS \cite{richard} and from 0.30 to 0.45 for MACS models \cite{richard_private2010}.   The selected clusters   are listed in table \ref{clusterreference}.

The modeling procedure is the same as described in section \ref{sec:abell1689}. 
To make a reasonable comparison,  the input source file  for each cluster should have the same number of   sources. Moreover, the whole source area should have the same size and be sufficiently sampled to cover the multiple image areas described by  the caustic  lines.


\subsection{Estimating $\beta$}

  \begin{figure}
 \centering  
\includegraphics    [scale = 1, bb=0 0 800 283]   {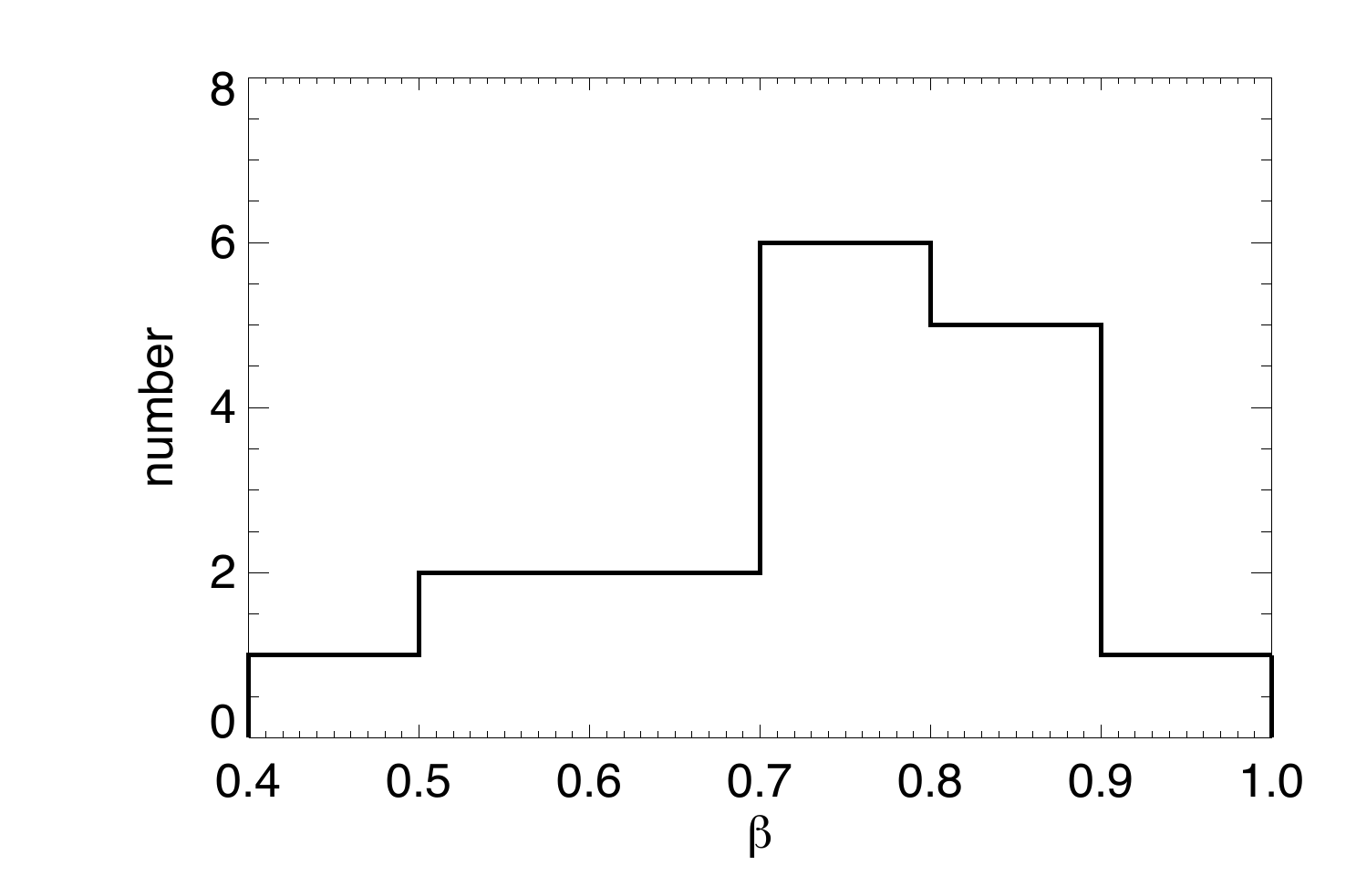}
\caption{  The   distribution of parameter $\beta$   for probability   distribution functions of 17 clusters. The modeling sources are at $z=3$.     The  probability   distribution functions are fitted to the modeling data of    time delays within 1000 days.   The parameter     range is   $\beta \in [0.41,  0.91]$.  }. 
  \label{slope}
\end{figure}
As for  Abell 1689 (section \ref{subsec:massanalys}), we   fit  power-law distribution functions to time-delay distributions generated from the  17 cluster models.  The   functions are fitted to the data with time delays less than 1000 days. The cluster masses    \cite{richard_private2011} \cite{richard}  and the   fitting values of $\beta$ are shown in table \ref{clusterreference}. 
 The distribution  of the parameter $\beta$ is shown in figure \ref{slope}.   The mean value is $\bar \beta= 0.75$,  
 and the median value is $\dot \beta= 0.79$.  
 With the least squares method, if the   clusters are weighted equally to each other,  we   determine that   the best-fitting slope   is   $\widetilde  \beta   = 0.77$  for the 17 clusters, with  standard deviations in $\log P(\Delta t)$ in the  range      [0.11, 0.30].  As a consequence, to a good approximation, $\beta$ can be fixed  in equation (\ref{function_betac}).   

 \subsection{Parameter estimation}

   \begin{figure}
\includegraphics  [scale = 0.5, bb= 0 0 800 920]  {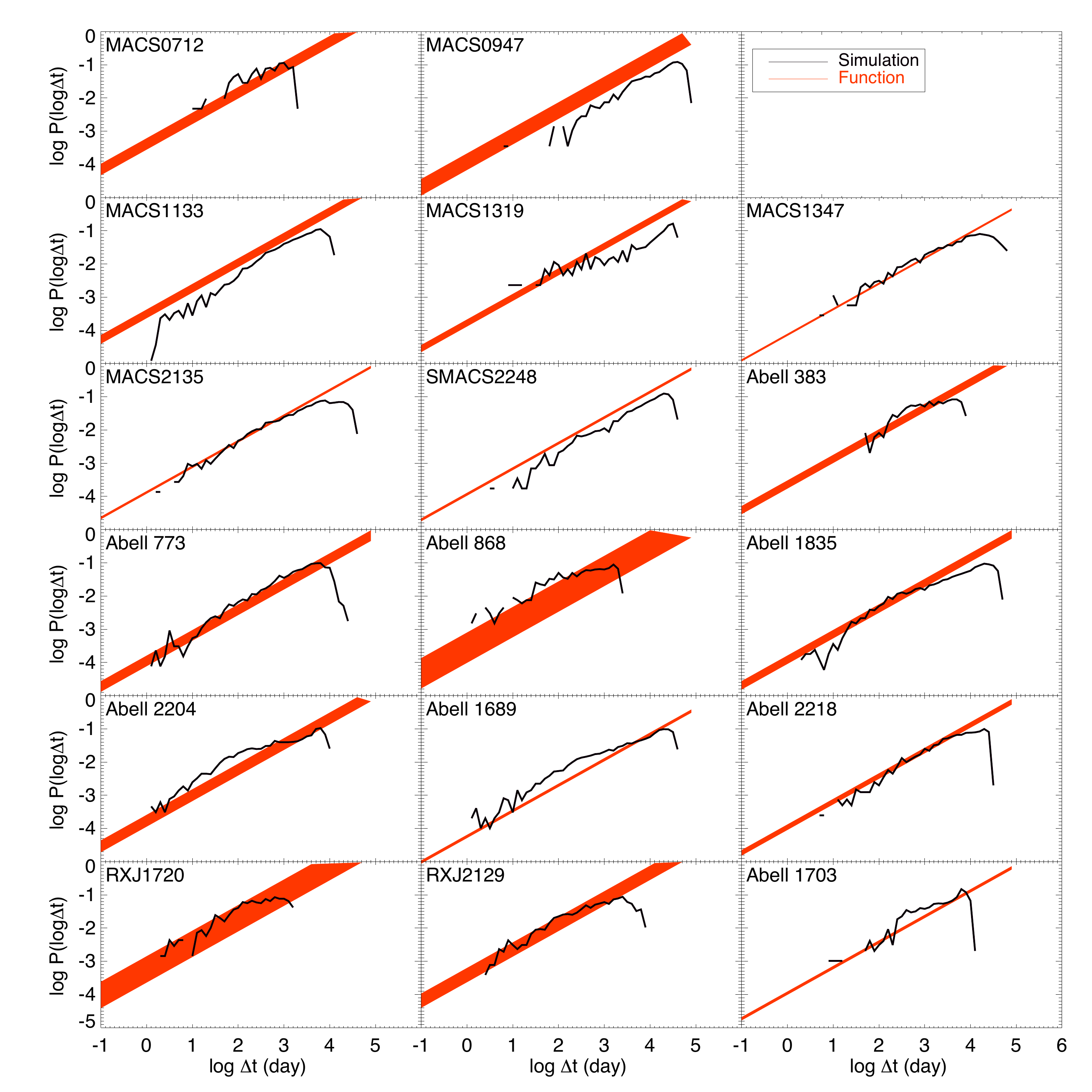}
\caption{ The logarithmic probability   distribution functions  for 17  clusters with background sources at $z=3$.  The solid curves represent the modeling results of time-delay distribution of 17 clusters. The   red shaded regions represent   the logarithmic probability distribution function from equation (\ref{function_parameter}).  The uncertainties of the distribution functions arise from the uncertainties in the mass.     For MACS1133,   the input source area is 10 $\times$  10 arcsec. For other  clusters,   the input source areas are  60 $\times$  60 arcsec. }  
  \label{function3}
\end{figure}

 We   fix   the  value of $ \widetilde  \beta = 0.77$ in the  logarithmic probability distribution function   (\ref{function_betac}) and fit the function to the simulated  data,   
and  then find the best-fitting value  for constant $C^{'}_{2}$. With smallest deviations,  we find   that the         best-fitting value    is      $C^{'}_{2} = -3.28$.  So the logarithmic probability distribution function (\ref{function_betac}) can  be    written as 
  \begin{equation} 
    \label{function_beta}
  \log P ( \log  \Delta t) =   \widetilde  \beta \log \Delta t  - 2     \widetilde  \beta \log  M_{250}  - 3.28,  
\end{equation} 
  or
  \begin{equation} 
    \label{function_powerlaw}
 P (\log  \Delta t) =  5.27 \times 10^{-4}  \Delta t^{ \widetilde \beta}  / M_{250} ^{2   \widetilde \beta}.  
\end{equation} 
If we introduce $  \widetilde  \beta = 0.77 $, then the logarithmic probability   distribution function is simply  
   \begin{equation} 
    \label{function_parameter}
\log P ( \log  \Delta t) =   0.77\log \Delta t  - 1.54 \log  M_{250} -3.28.   
\end{equation}
For a cluster with $M_{250} = 2 \times 10^{14}  \  \textrm{M}_{\odot}$, the probability of a time delay less than 1000 days   is about $0.025$.  
In figure \ref{function3}, we present  modeled time-delay distributions   for  all   17 clusters and their logarithmic probability distribution functions      (\ref{function_parameter})  with sources at $z=3$.     Among the 17 systems,  MACS1133 has much smaller multiple-image  area  enclosed by the caustic lines. To make the small time delays    detectable,  for MACS1133, we change the input source area to 10 $\times$  10 arcsec. For other  clusters, we keep the input source as  60 $\times$  60 arcsec.

\section{The rate of lensed supernovae in Abell 1689}
\label{sec:snr}



With the help of the gravitational lensing amplification, we can potentially observe   supernovae  which would otherwise be too faint for detection. If a source   is located inside the multiply imaged surface defined by the  caustic line, two or more   images will be generated.  
We may therefore observe   multiple images of a supernova  in the source galaxies.

For Abell 1689, we know 35 multiply imaged sources and   116 corresponding  multiple images.  For each multiple-image pair producing a time delay,  
we call the   image arriving  first to the observer   the {\it leading image}.  For example, for a source with 3 multiple images: image1, image2, image3,  assuming the images have arrival times $\tau_{image1}  <  \tau_{image2}  < \tau_{image3}$,   there are three image pairs: $\textrm{pair} 12$,  $\textrm{pair}13$ and  $\textrm{pair}23$.  The corresponding leading images are $\textrm{image}1$, $\textrm{image}1$ and $\textrm{image}2$, accordingly.  So in this three-image system, only the image with the longest arrival time (in this case, it is image3) cannot be the leading image. That is,  for 1 source having 3 multiple  images, the number of the leading images are $3-1 = 2$.  Therefore,  in a lensing   system with $m$ sources and $n$ corresponding multiple images, the number of the leading images are $n -m$.  
According to this definition,  in Abell 1689,  there are $(116-35 = )$   81   leading images in total.

To estimate the probability of observing a leading supernovae image in  Abell 1689,  we need to know the supernova  rate. 
The   models for describing the supernova rate are dependent on the types of the supernova.    
 The rate of core-collapse supernovae ($\textrm{SNR}_\textrm{cc}$)  can be obtained from the   star-formation rate:

 \begin{equation} 
   \label{abmode_sfm_cc} 
  \left(\frac{\textrm{SNR}_\textrm{cc}} {  \textrm{yr}^{-1} }  \right) =  k_\textrm{cc} \cdot 10^{-3}      \left(  \frac{\textrm{SFR}} { \textrm{M}_{\odot} \  \textrm{yr}^{-1} }\right),
\end{equation} 
where the parameter  $k_\textrm{cc}$ can be determined by measuring the  $\textrm{SNR}_\textrm{cc} $ and $\textrm{SFR}$.  Here the factor of   $10^{-3}$  is multiplied into the function to simplify the parameter  $k_\textrm{cc}$.  We also multiply factors in the following functions    (\ref{abmode}) (\ref{abmode_sfm})  for the same reason.    The   $\textrm{SNR}_\textrm{cc} $ and $\textrm{SFR}$ can be derived from observational data \cite{dahlen_19}  \cite{Giavalisco_2004}. 
By using the  core-collapse SN rate density and comparing it against the
SFR density, the parameter is constrained to   be
  $k_\textrm{cc} =  7.5 \pm 2.5  $  \cite{Scannapieco_2005}.    Alternatively, by using a Salpeter IMF and  
a progenitor mass ranging  between 8 and 50 solar masses, the parameter is estimated to be 
 $k_\textrm{cc} =   7 $      \cite{Riehm}.   
   In this paper, we choose  $k_\textrm{cc} = 10  $ as the upper limit, and $k_\textrm{cc} =  5  $ as the lower limit.


For   Type Ia supernovae, we use    the popular   
  \textquotedblleft two-component\textquotedblright  \    model to estimate the Type Ia supernova  rate ($\textrm{SNR}_\textrm{Ia}$) \cite{Scannapieco_2005}       
   \cite{Mannucci_2005}  \cite{Sullivan_2006}:
   \begin{equation} 
   \label{abmode} 
       \left( \frac{\textrm{SNR}_\textrm{Ia}} { \textrm{yr}^{-1} }\right)  = \hat A \cdot 10^{-10}  \left( \frac{M_{\star}}{  \textrm{M}_{\odot} } \right)^{\alpha}   +\hat  B  \cdot 10^{-3}  \left(  \frac{ \textrm{SFR}}{  \textrm{M}_{\odot}   \textrm{yr}^{-1} }\right),
\end{equation} 
where $M_{\star}$ is the host  stellar mass,   
 and $\alpha$ denotes the exponent of the  stellar  mass.     The first component describes the stellar mass contribution. The second component describes the host galaxy star-formation contribution. 
  For parameters $ \hat A$ and $\alpha$, we choose $ \hat A =  1.05 \pm 0.16  $ and $\alpha = 0.68 \pm 0.01$ \cite{smith_2011}.  
 The parameter $ \hat B$ can be related to the 
 $\textrm{SNR}_\textrm{cc} - \textrm{SFR}$  relation (\ref{abmode_sfm_cc}),   
  \begin{equation} 
   \label{beta_k} 
    \hat  B = k_\textrm{cc} \Theta, 
 \end{equation} 
where     $\Theta = \textrm{SNR}_\textrm{Ia} / \textrm{SNR}_\textrm{cc}$. The value of $\Theta$ has been estimated at redshift up to $z  \sim 1.5$ \cite{dahlen_19}.  At redshift $z < 1$, the ratio of $\textrm{SNR}_\textrm{Ia} / \textrm{SNR}_\textrm{cc}$   approximately ranges between $\Theta = 1/2$  and $\Theta = 1/4$,  which is consistent with the result  $\Theta =  0.35 \pm 0.08$ in nearby galaxies \cite{Mannucci_2005}.  At higher redshift   $z > 1.0$, 
inspired from  figure 3 of \cite{dahlen_19},  we assume   
\begin{equation}
 \label{theta_z} 
 \Theta =    \frac{1}{15};     \qquad   1.0 \leq z.
\end{equation}
 Considering all sources in Abell 1689 have redshifts $z >1.0$, in this work,  $\hat B = 0.5 \pm 0.17$.  
This value is consistent with  $\hat B = 0.39 \pm 0.07$ (based on redshift $0.2 < z < 0.75$ \cite{Sullivan_2006}).

 From their magnitude in F775W, we   estimate  the flux and the luminosity and   then constrain their SFR  \cite{robert} as
  
     \begin{equation} 
   \label{sfr_snr} 
\left(   \frac{ \textrm{SFR} }{ \textrm{M}_{\odot} \: \textrm{yr}^{-1} } \right) =     1.4 \times 10^{-28} \left(    \frac{\bar{L}_{\nu}} { \textrm{erg} \, \textrm{ s}^{-1} \, \textrm{Hz}^{-1} }\right) .
\end{equation} 
This conversion between UV flux and the SFR is for rest wavelengths, ranging from 1500 $\textrm{\AA}$  to 2800 $\textrm{\AA}$, while our data \cite{MarceauLimousin} have observed wavelength in the range 6900 $\textrm{\AA}$ to 8600 $\textrm{\AA}$.  We assume these galaxies have flat spectra, as is typical for star-forming galaxies, so 
 we can calculate the luminosity from the flux. 

  \begin{figure}

 \centering

\includegraphics  [scale = 1, bb=0 0 800 283] {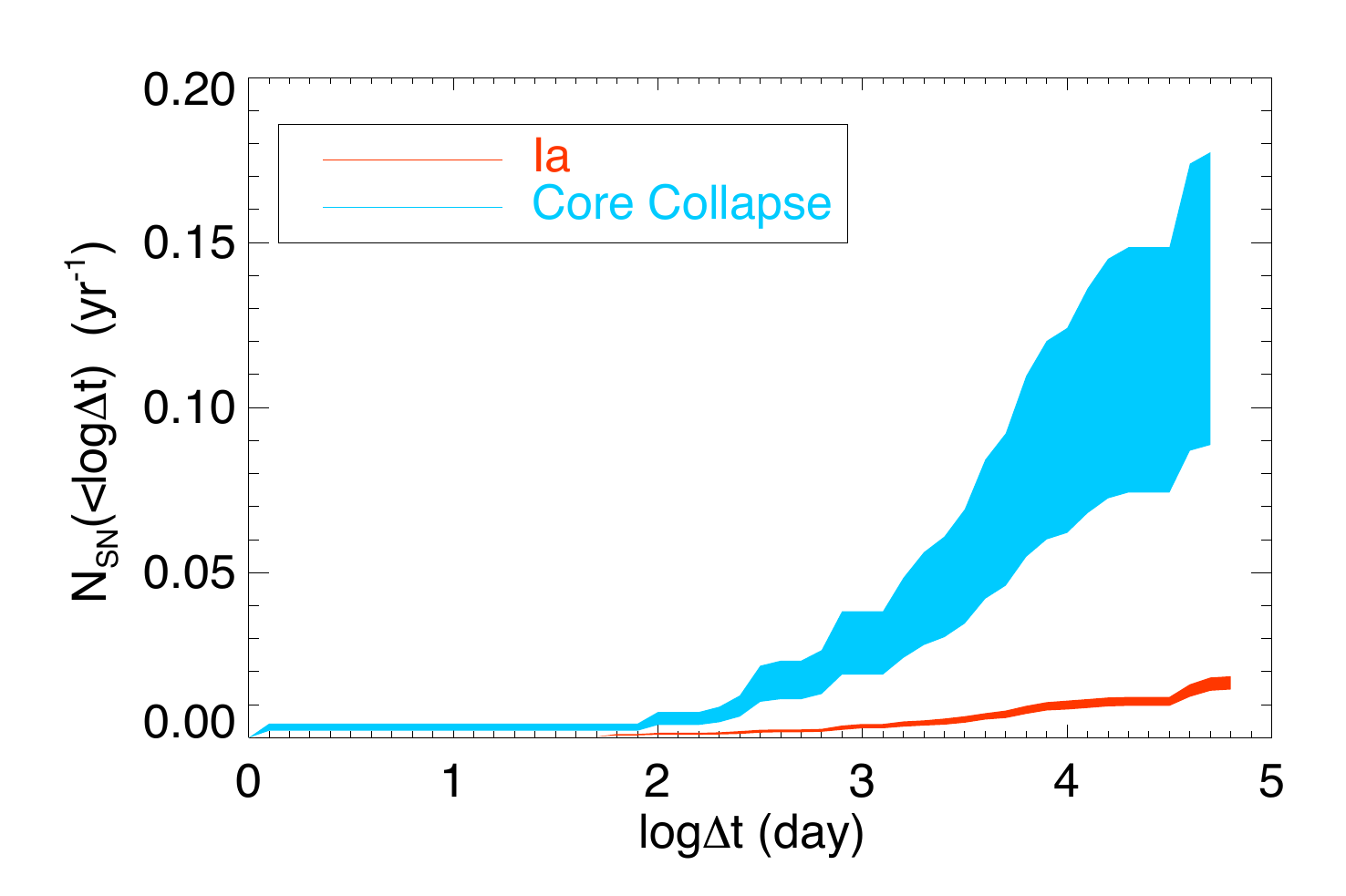}
\caption{ The   cumulative rate ($N_\textrm{SN}$) of   the leading   images  of lensed supernovae in  Abell 1689, derived from equations (\ref{abmode_sfm_cc})  and  (\ref{abmode_sfm}).    The magnitude threshold  is  26.5 and   a total number of 70   among the 72 leading images  are included  because of the magnitude threshold.   The time delays are calculated based on the modeled source positions.     The red curve represents the estimated cumulative  $\textrm{SNR}_\textrm{Ia}$, while the cumulative $\textrm{SNR}_\textrm{cc}$ is plotted in blue curve.   The uncertainties of $N_\textrm{SN}$ arise from the upper and lower limits of the parameters in the functions (\ref{abmode_sfm_cc})  (\ref{abmode})  (\ref{abmode_sfm}). 
}
  \label{supernova}
\end{figure}
 
Note that  multiple images from the same source  should have the same inferred luminosity.  
 To infer the luminosity of each image, we need to know their fluxes.  With magnitudes of 116 images \cite{MarceauLimousin} \cite{richard_private2011},   we can estimate their  fluxes.   Using {\it Lenstool},   we can  get a magnification map on the image plane then read the values of the magnification on the map.      Considering the gravitational magnification effect and {\it k correction}                           \cite{Hogg_2002}   \cite{van_Dokkum}, the flux can be calculated as  
      \begin{equation} 
   \label{mag_snr} 
 \log (F_{\nu}) = [m_ \textrm{AB} + 2.5 \log ( |\mu|) + 2.5 \log (1+ z) + 48.6] / (-2.5),
\end{equation} 
 where $\mu$ is the gravitational magnification factor, and $z$ is the source redshift.   
When  images are located close to the caustics,   their   gravitational magnification factors ($\mu$) may be very large.  Thus, the  values of   fluxes of these images derived from equation \ref{mag_snr} are uncertain, and their  luminosities estimated from the fluxes may not be reliable.  So we  neglect these images and average other luminosity values of images of the same source  to constrain $\bar{L}_{\nu}$.  When we know their luminosities, from equation (\ref{sfr_snr}), we can calculate their SFR,  and then   $\textrm{SNR}_\textrm{Ia}$ (\ref{abmode_sfm})  and $\textrm{SNR}_\textrm{cc}$ (\ref{abmode_sfm_cc}).

 We also need to calculate the stellar masses   and the star-formation rates for each galaxy.
  To estimate the stellar-mass contribution to the SNR,  we separate the images into two groups. In group one, with data of   photometry from {\it HST}/ACS in bands B, V, I, Z, ground-based near-infrared imaging and {\it Spitzer}/IRAC photometry  \cite{richard_private2011},  we derive their stellar mass from SED fitting \cite{Walcher_2011}.   In group two with insufficient photometric data, we  estimate the mass contribution based on   the ratio  of mass  and star-formation contribution to the SNR derived from  group one.     
   The median value of this ratio, is only   2.9\% (4.2\%) of the upper (lower) limits of the star-formation part.  In this paper, we choose its median value and assume   the mass part contributes  3.5 \%  
 of the star-formation part. Therefore, for group two, the $ \textrm{SNR}_\textrm{Ia}$  may be estimated   as 
 \begin{equation}
   \label{abmode_sfm} 
       \left(   \frac{\textrm{SNR}_\textrm{Ia}} { \textrm{yr}^{-1} }\right)  = 1.035 \cdot  \hat  B  \cdot 10^{-3}  \left(  \frac{ \textrm{SFR}}{\textrm{M}_{\odot}    \textrm{yr}^{-1} }\right).   
        \end{equation}

%
 We   assume  Type Ia supernovae have  absolute magnitude $ M = -19.3 $  mag \cite{dahlen_19}, and core-collapse supernovae  have $ M = -17.0 $ mag  \cite{DAndrea}.     With their absolute magnitudes, considering the {\it k correction}, we calculate the apparent  magnitudes for each supernova.  
 Their     apparent magnitudes of a supernova    can be expressed as:
  \begin{equation}  
      m  = M + 5 \log_{10}( D_L /10 \textrm{ pc}) - 2.5 \log_{10}(|\mu|) -2.5 \log_{10}(1+z).
      \label{magthres}
\end{equation} 
 Here $D_L$ denotes the luminosity distance.  
This equation is  used to   constrain  the magnitude when considering the magnitude thresholds in figures   \ref{supernova}  and  \ref{supernova_threshold}.

The cumulative rate of observing the leading images of the lensed supernovae in Abell 1689 is shown in figure \ref{supernova}.  Among the total 81 leading images, we exclude 14 images from 5 sources with insufficient  magnitude data   \cite{MarceauLimousin},   so there are 72 leading images to be considered.     Here   $N_\textrm{SN} (< \log \Delta t)$ represents the cumulative rate of the supernova images observable in one year.  We set the detection threshold to 26.5 mag in all bands.   In that case,      a total number of 70   of the 72 leading images  are included.         The resulting probabilities are   $0.004 \pm 0.002$ for the Type Ia supernovae and $0.029 \pm 0.001$ for the core-collapse supernovae in one year, assuming time delays  less than 1000 days.

 \begin{figure}

 \centering

\includegraphics  [scale = 1, bb=0 0 800 283] {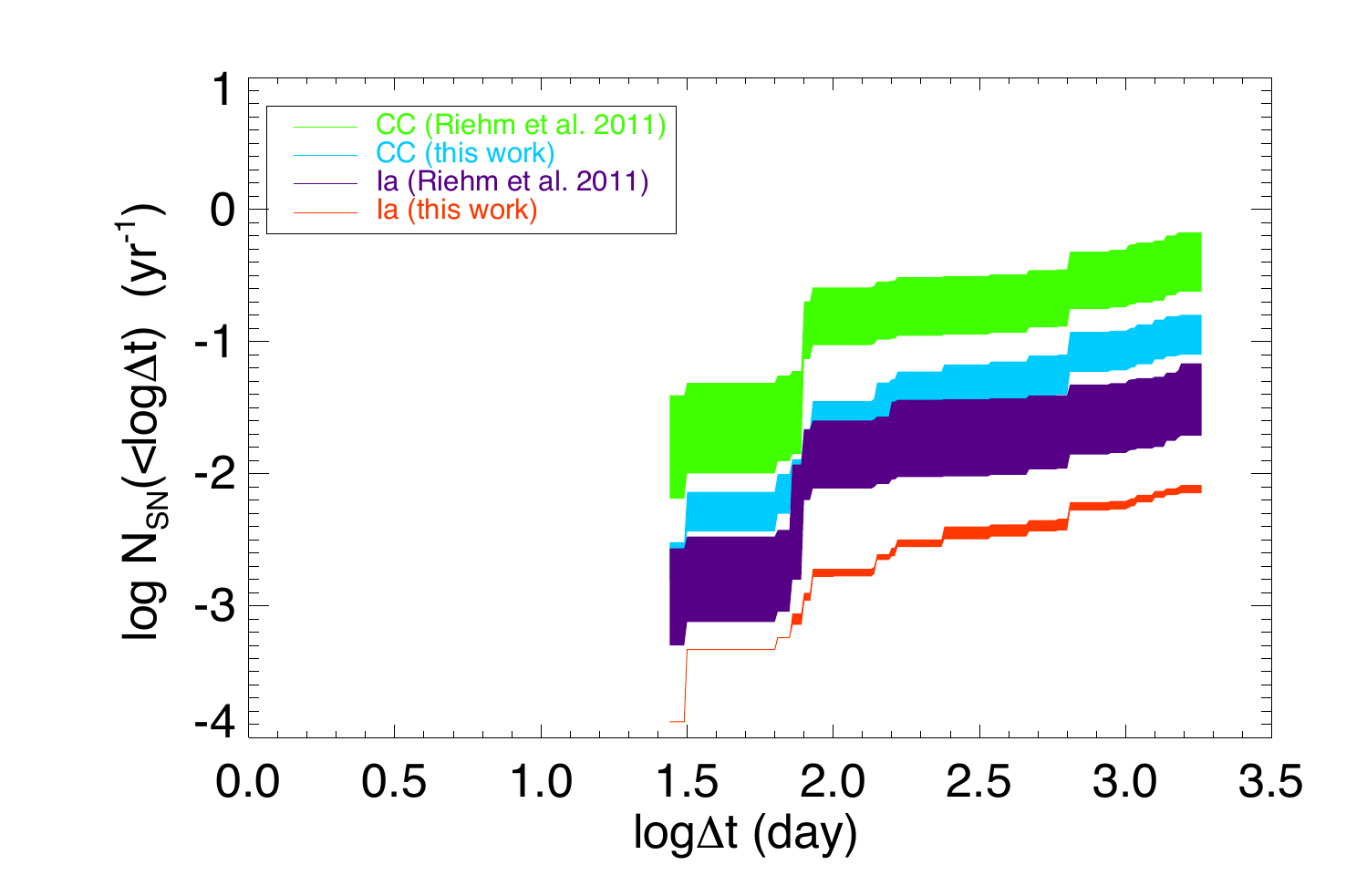}
\caption{ The comparison of the supernova  rate estimated from functions (\ref{abmode_sfm_cc}) and (\ref{abmode_sfm}), and the supernova  rate from  \cite{Riehm},  with time delays less than 5 years. The    green and purple curves represent the logarithmic  cumulative $ \textrm{SNR}_ \textrm{cc}$ and $ \textrm{SNR}_ \textrm{Ia}$ from \cite{Riehm}. Our results are shown in blue   for $\textrm{SNR}_\textrm{cc}$ and in  red   for $\textrm{SNR}_ \textrm{Ia}$. The uncertainties of $\log N_ {SN}$ arise from the upper and lower limits of the parameters in the functions (\ref{abmode_sfm_cc})  (\ref{abmode})  (\ref{abmode_sfm}).   At $\log \Delta t \sim 1.8$, there is a fast increase in the  $\log N_\textrm{SN}$.    Firstly,   in this region, there are more leading images included, so both our results and the results from  \cite{Riehm}  raise quickly. Secondly, the leading images       have much larger SNR and uncertainties in their result. This causes an even faster raise and higher upper limit in their curves.  
 }
  \label{hersnr}
\end{figure}

 We also compare our results to the results from other groups \cite{Riehm}. The result is shown in figure  \ref{hersnr}. 
The difference in   supernovae rates may arise from the difference in the lens model and/or the SNR prescription. For example, different lens models may produce different magnification factors for each images, and then produce different luminosities, and SFR. 
In addition, the different parameters chosen in the SNR model (\ref{abmode}), e.g., $k_\textrm{cc}$, $ \hat  A$, $\alpha$, $ \hat  B$ may also affect the constraint on SNR. In this paper, we choose  $k_\textrm{cc} = 7.5\pm2.5$, $ \hat  A= 1.05 \pm 0.16$, $\alpha =    0.68 \pm 0.01$, $ \hat  B = 0.5 \pm 0.17$, while in \cite{Riehm}, the parameters are $k_\textrm{cc} = 7.0$, $ \hat  A= 4.4^{+1.6}_{-1.4}$, $\alpha =    1.0$, $ \hat  B = 2.6 \pm 1.1$.  Larger values of $\hat A$, $\alpha$, $\hat B$ chosen will cause a higher   SNR estimate.  
We applied the  parameters  of \cite{Riehm} to the functions  (\ref{abmode_sfm_cc})  (\ref{abmode})  (\ref{abmode_sfm}), in an attempt  to reproduce their results. 
The results fit well to their   $ \textrm{SNR}_\textrm{Ia}$, but a significant  discrepancies remain for  $ \textrm{SNR}_\textrm{cc}$.  

We   estimate the probability of observing the leading supernova  images as a function of  magnitude threshold in figure \ref{supernova_threshold}.    
%
The figure shows the cumulative rates of  observable leading supernova images as a function of  magnitude threshold, with time delays   less than 1000 days.  For a magnitude threshold of  27.0,  we can observe  
 $0.044 \pm 0.015$  core-collapse supernovae per year, with time delays less than 1000 days.  Under the conditions of small time-delay scales and limited magnitude threshold,  the probability of observing a leading supernova image  is quite low. 

 \begin{figure}

 \centering

\includegraphics  [scale = 1, bb=0 0 800 283]  {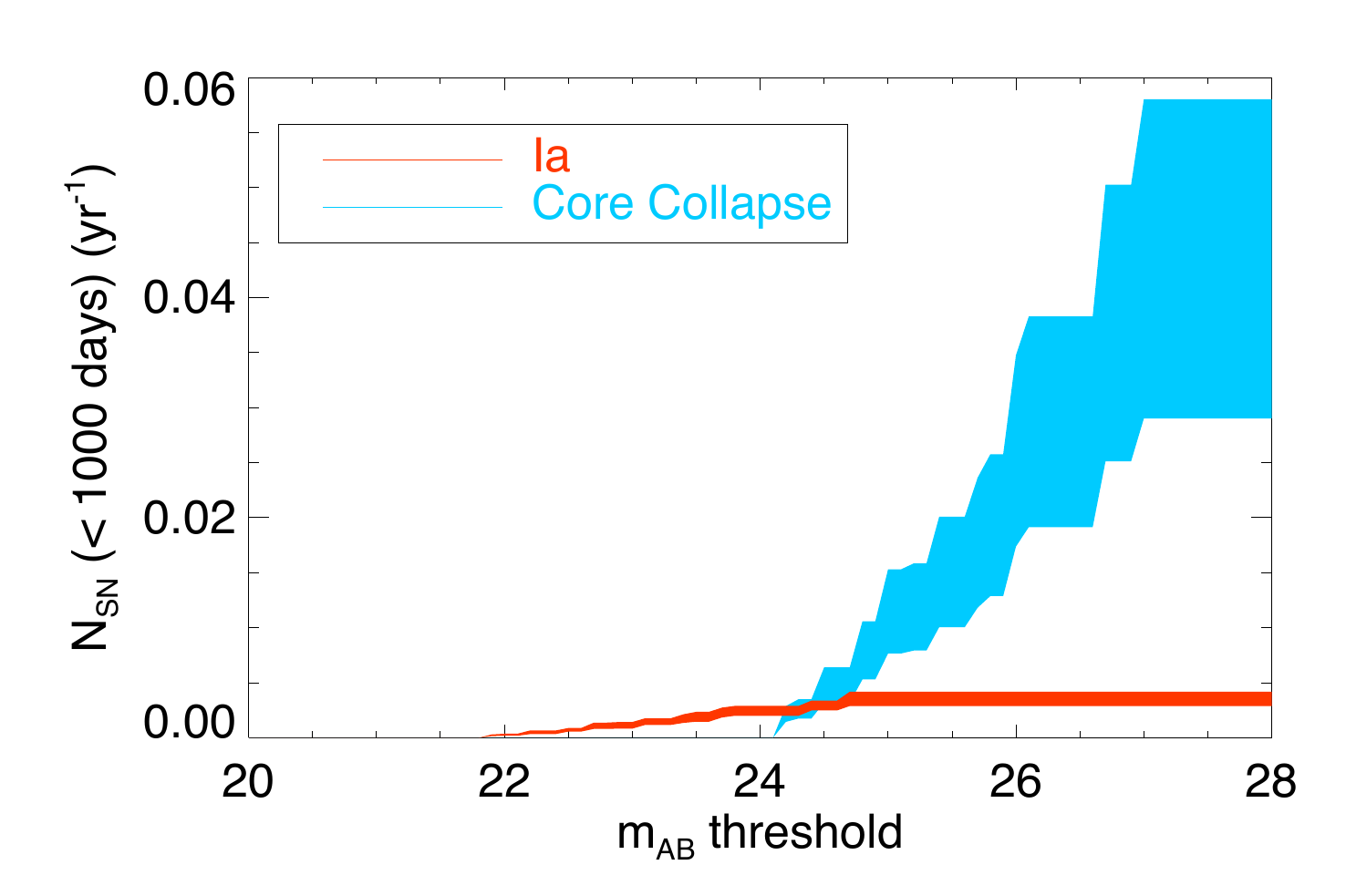}
\caption{The cumulative rate of   observing a  leading supernovae image, as a function of  magnitude threshold. The images have        time-delay separations less than 1000 days.  The uncertainties of $N_\textrm{SN}$ arise from the upper and lower limits of the parameters in the functions (\ref{abmode_sfm_cc})  (\ref{abmode})  (\ref{abmode_sfm}). 
The  Type Ia supernovae involved are much brighter than those of the core-collapse supernovae. This causes a `shift' in the distributions of the rates of Type Ia supernovae.  At  $m_\textrm{AB}$ around 24.7 for Type Ia supernovae, the curve flattens out. This is because all the Type Ia supernovae involved in the calculation have magnitudes smaller than about 24.7.   
The {\it HST} telescope has the magnitude limit in detecting the lensed sources. The {\it James Webb Space Telescope} may   detect   fainter lensed sources behind Abell 1689. As a sequence,    Type Ia supernovae fainter than 24.7 will be detected,  and the cumulative rate of observing a leading Type Ia supernovae ($N_\textrm{SN}$) will  increase at  $m_\textrm{AB} > 24.7$  as well. This is also the case for the  core-collapse supernovae when $m_\textrm{AB} > $ 27.0.  
}
  \label{supernova_threshold}
\end{figure}

\section{Summary and discussion}
\label{sec:Summaryanddiscussion}

We analyzed  the  time-delay distributions in strong lensing systems.  
We found   that we can describe the probability  distribution of time delays as a power law function (\ref{function_betac}).      In the function, there are two parameters, $M_{250}$,  $\beta$,  and a constant   $ C^{'}_{2}$.   Modeling  with  mass profiles of  SIS, NFW and dPIE (in figure \ref{theorysispiemdnfw}), we found  that the parameter $\beta$ is strongly affected by the slopes of the mass profiles of the lensing clusters. The shallower the inner density profile and the steeper the outer density profile are,    the more the   time-delay distribution will be stretched out to both the  higher and the lower end, causing a lower $\beta$. By modeling   Abell 1689, we found   that the massive galaxies and halos mainly produce large time delays, while small time delays are predominated produced by substructures (galaxies) in the cluster.    We also simulated and verified that the time-delay distribution generated by `real' clusters with more than 2 multiple images from the same sources also obey the power-law distribution  in figure  \ref{abell1689function_1}.   

 To estimate the   parameter and the constant     in the logarithmic probability   distribution function,  we modeled 17 strong lensing   clusters as shown in figure \ref{function3}, using their well calibrated mass models.     With  the fixed  best-fitting slope $\widetilde  \beta = 0.77$, we determined the best-fitting value  of   $ C^{'}_{2}$ to the function (\ref{function_betac}).       
The resultant logarithmic probability   distribution function  (\ref{function_parameter}) enables us to estimate the time-delay distribution of a cluster with known  mass.    

 We also calculated the probability of  observing the leading  images of the lensed supernovae in Abell 1689.  
 The   $\textrm{SNR}_\textrm{cc}$ can be derived from the SFR (\ref{abmode_sfm_cc}).    
   The \textquotedblleft  two component\textquotedblright  \  model   was applied to constrain the $\textrm{SNR}_\textrm{Ia}$.  We constrained the parameters in the function (\ref{abmode}), and calculated the SNR for Type Ia supernova.  
 We estimated the luminosity from magnitudes of images in Abell 1689 (\ref{mag_snr}), derived the SFR from the luminosity (\ref{sfr_snr}), and then   estimated the probability of observing a leading  supernova image 
   in the system as shown in figure \ref{supernova}.   Considering a typical magnitude limit of observations with  $m_ \textrm{AB} = 26.5$,  we can observe $0.004 \pm 0.002$ Type Ia supernovae and $0.029 \pm 0.001$   core-collapse supernovae  per year.  We compared the  results in this work to   \cite{Riehm} as shown in figure \ref {hersnr}, and discussed the possible reasons which may cause the differences.
  
 We also constrained the cumulative rate of   observing a  leading supernovae image, as a function of the  magnitude threshold (in figure \ref{supernova_threshold}). 
        If  the magnitude limit is lowered to 27.0,  the probability of  observing   the leading images of the  core-collapse supernovae will be up to $0.044 \pm 0.015$ per year,   with image separations within 1000 days.  This probability is quite low, which means that detecting time delays from lensed supernovae will be challenging with current facilities.

\acknowledgments
 \label{Acknowledgments}

  We   thank Johan Samsing for discussions on his work on time-delay distributions. We thank Danuta Paraficz and  {\'A}rd{\'{\i}}s El{\'{\i}}asd{\'o}ttir for many helpful discussions on gravitational lensing. We also   thank Claudio Grillo, Andrew Zirm  and Teddy Frederiksen  for their helpful comments on the paper,    and Enrico Ramirez-Ruiz for   discussion on Type Ia supernova progenitor models.    The Dark Cosmology Centre is funded by the Danish National Research Foundation.

\bibliographystyle{JHEP} 
\bibliography{ref}

 \end{document}